\documentclass[dateno]{article}

\usepackage{arxiv}

\usepackage[utf8]{inputenc} 
\usepackage[T1]{fontenc}    
\usepackage{hyperref}       
\usepackage{url}            
\usepackage{booktabs}       
\usepackage{amsfonts}       
\usepackage{nicefrac}       
\usepackage{microtype}      
\usepackage{lipsum}
\usepackage{graphicx}
\graphicspath{ {./images/} }

\title{Envision of mmWave Wireless Communication with Artificial Intelligence}

\author{
 Quanda Zhang \\
  School of Telecommunication\\
  Daqing Normal College\\
  \texttt{quandazhang2002@dqnc.edu.cn} \\
   \And
 Hudi Wang \\
  School of Telecommunication\\
  Daqing Normal College\\
  \texttt{hdwang@dqnc.edu.cn} \\
}

\begin{document}
\maketitle

\begin{abstract}
The future wireless communication system faces the bottleneck of the shortage of traditional spectrum resources and the explosive growth of the demand for wireless services.
Millimeter-wave communication with spectral resources has become an effective choice for the next generation of wireless broadband cellular communication. However, the transmission path loss is large and oxygen and water molecules absorb
Characteristics such as seriousness have brought great challenges to millimeter wave communication, and it is necessary to seek a technical approach different from low-frequency wireless communication. In the analysis of millimeter wave transmission characteristics
After the analysis, the research progress of millimeter wave communication technology and the RF front-end is comprehensively analyzed, and the technology of millimeter wave communication is thoroughly analyzed with technical challenges and proposed corresponding research directions. Finally, the advances of artificial intelligence and machine learning are also applied into millimeter wave communication, so we also cover these parts in this paper.
\end{abstract}

\section{Introduction}
With the rapid development of electronic information technology, especially wireless
With the development of communication technology and the popularization of 4G wireless communication technology, in just 10 years, the amount of IP data processed by wireless networks has exceeded more than 100x: from 2010 to less than 3 EB (Exabyte, Exabyte bytes) to 190 exabytes in 2018, and even on track for 2020 More than 500 exabytes. Video data processing is by far the main factor behind the growth of data processing in wireless networks. Adaptive Personal Communication Technology. Emerging Mobile Internet and Internet of Things Technologies are also advancing by leaps and bounds.
In the future, the number of network devices and smart terminals continues to grow,
the demand for communication services is getting higher and higher, resulting in the demand for network traffic exponential growth.

In order to meet the growth of communication traffic thousands of times in the next few years,
development needs, and research on wireless communication technology in industry and academia
People are in full swing on key technologies for 5G mobile communications
research~\cite{shafique2020internet}. The key technology of 5G mobile communication is embodied primarily in wireless transmission
technology and network technology~\cite{alkhateeb2014channel,el2013practical}; the former mainly includes large-scale
line array, ultradense networking, new multiple access and full-spectrum access, etc.~\cite{zhang2014dynamic}.
The latter mainly include self-organizing network (SON), soft-defined network
network (SDN), network function virtualization (NFV), etc.~\cite{gong2013virtual, zhu2012dynamic,yin2013spectral,xia2014survey,kirkpatrick2013software,lu2017game}.
The main development goal of mobile communication technology is to communicate with other wireless mobile
devices. Communication technology is closely linked, providing for the rapid development of the mobile Internet.
Provide ubiquitous basic business capabilities, mainly from three aspects
Carry out research and exploration: first, through the introduction of new wireless transmission technology~\cite{zhang2019understanding}
The resource utilization rate is increased by more than 10 times on the basis of 4G;
second, by introducing a new architecture (such as ultra-dense networking, control
noodle making and data surface separation, etc.) and deeper intelligent capabilities
Improve the throughput of the whole system by about 25 times; the final further
mining new frequency resources (millimeter wave, visible light, laser communication,
etc.), so that the mobile communication frequency resource is expanded by about 4 times~\cite{el2012low,el2012capacity,khan2011mmwave,dhillon2014fundamentals,rajagopal2011antenna}.

Millimeter-wave (mmWave) wireless communication is a key technology that enables the high data rates required for 5G communication. These high frequencies, which range from 30 to 300 GHz, allow for the transmission of large amounts of data over short distances, making them ideal for high-speed, high-capacity applications such as 5G.

One of the main advantages of mmWave technology is its ability to support massive amounts of data transmission. This is because the high frequencies used in mmWave communication allow for a wider bandwidth, which means that more data can be transmitted in a given amount of time. This increased capacity is critical for 5G, which is expected to support a wide range of applications, from enhanced mobile broadband to the Internet of Things (IoT) and beyond.

Another important benefit of mmWave technology is its ability to support ultra-low latency communication. Latency is the time it takes for a signal to travel from one point to another, and it is a crucial factor in the performance of any wireless communication system. In mmWave systems, the short distances over which the signals travel and the high frequencies used allow for ultra-low latency communication, making it possible to support real-time applications such as augmented and virtual reality.

In addition to its high data rates and ultra-low latency, mmWave technology also offers several other benefits for 5G communication. For example, mmWave signals are highly directional, which means that they can be focused on specific areas, reducing interference and improving signal strength. This is especially useful in urban environments, where there are many potential sources of interference. Additionally, mmWave technology can support multiple-input multiple-output (MIMO) communication, which uses multiple antennas to transmit and receive signals simultaneously, further increasing capacity and reliability.

Overall, mmWave technology is an essential component of 5G communication, offering high data rates, ultra-low latency, and other key benefits that make it possible to support a wide range of applications and services. As 5G continues to develop and expand, mmWave technology will play a critical role in enabling the high-speed, high-capacity communication that is required to support the growing demand for data.

\section{Challenges}
\label{sec:headings}

According to Shannon's Theorem, it is known that expanding the channel capacity is the most direct way.
The most effective way is to increase the system bandwidth. Currently used for mobile communication, the bandwidth of the microwave band is only 600 MHz, and 5G mobile in 2020 requires about 1 GHz of bandwidth, so one needs to dig
digging for new spectrum resources to meet the needs of mobile Internet traffic in the next 10 years, 1,000 times growth in volume. Using the millimeter wave frequency band
Potentially large bandwidth (30~300 GHz) to provide higher data rates
However, the millimeter wave~\cite{bhartia1984millimeter,rappaport2013millimeter,pi2011introduction,marcus2005millimeter,mishra2019toward,xiao2021survey,lu2017convolutional} has fairly rich spectrum resources, which can be utilized for other communication purposes.
For example, it can be used for realizing high-speed wireless transmission
output, but many technical challenges are encountered in the process of implementing mmWave cellular network communication many technical challenges:

\begin{itemize}
    \item Shadowing and Rapid Channel Fluctuations: Shadowing effects can cause an outage or poor mmWave propagation channel quality. Meanwhile, the security of the communication channels is also very important, so there have been works, which use encryption techniques~\cite{jin2018multi,goldsmith1994error} to protect the security of it. 
The linear relationship of the rate means that in the mmWave channel phase
dry times are very small, e.g. for a moving speed of 60 km/h
In terms of degrees, the Doppler spread at 60 GHz is about 3 kHz, that is, millimeter wave channels vary in milliseconds
characteristics, so how to quickly track mmWave channel changes.
It is an important problem in millimeter wave communication.

\item Multiuser coordination: research and application of millimeter wave communication
Focus on point-to-point links (such as cellular backhaul) or limit
LANs and personal area networks that limit the number of users and prohibit the
MAC protocol for multi-user transmission. 
For spatial multiplexing with spectral efficiency,
a new mechanism needs to be studied to coordinate the mmWave spatial multiplexing transmission
mechanism.

\item Power Consumption: Power consumption is linear with the sample rate, while
it is exponentially related to the number of sampling bits~\cite{carroll2010analysis}. For low power
consumption and low-cost equipment, on a large bandwidth and large scale, achieving high-resolution quantization on antennas is not practical; especially
in terms of high-definition video transmission, millimeter wave communication is facing
more severe technical challenges.

\item Range and Directional Communication: According to Friis'
law of propagation, the loss of an omnidirectional path in free space is proportional to the square of the frequency. Therefore, millimeter wave.
The communication system requires the use of large-scale array antenna communication technology
technologies such as beamforming and directional transmission to compensate for millimeter
wave propagation path loss.
\end{itemize}

In detail, we extend some of the above challenges and present more details and solutions to them.

One of the main challenges of millimeter wave (mmWave) wireless communication is its limited range. Because mmWave signals are highly directional, they have a limited range and are susceptible to interference from obstacles such as buildings and trees. This can make it difficult to establish and maintain a stable mmWave connection, particularly in urban environments where there are many potential sources of interference.

To overcome this challenge, several solutions have been proposed, including the use of beamforming technology. Beamforming uses multiple antennas to focus the mmWave signal in a specific direction, increasing its range, and reducing interference. Additionally, adaptive beamforming algorithms can be used to automatically adjust the beamforming pattern in response to changes in the environment, further improving the performance of the mmWave system.

Another challenge of mmWave technology is its sensitivity to weather conditions. Because mmWave signals are attenuated more easily than lower-frequency signals, they can be significantly affected by rain, snow, and other weather conditions. This can result in reduced signal strength and reduced capacity, making it difficult to maintain a stable mmWave connection in adverse weather conditions.

To address this challenge, several solutions have been proposed, including the use of advanced modulation schemes and error correction codes. By using more advanced modulation schemes and error correction codes, it is possible to increase the robustness of the mmWave signal, allowing it to maintain its performance in adverse weather conditions. Additionally, the use of multiple-input multiple-output (MIMO) communication can also help to improve the robustness of the mmWave signal by using multiple antennas to transmit and receive signals simultaneously.

In addition to these challenges, mmWave technology is also faced with the challenge of high power consumption. Because mmWave signals have a high bandwidth, they require a large amount of power to transmit and receive. This can be a significant challenge for mobile devices, which have limited battery life and are not always able to support the high power requirements of mmWave technology.

To overcome this challenge, several solutions have been proposed, including the use of power-efficient hardware and software design. Using power-efficient hardware and software, it is possible to reduce the power consumption of mmWave systems, making it possible to support mmWave communication on mobile devices with limited battery life. Additionally, the use of advanced power management techniques, such as dynamic voltage and frequency scaling, can also help reduce the power consumption of mmWave systems.

\section{Potential Opportunities}

This high-frequency communication technology offers many benefits, including high data rates, ultra-low latency, and the ability to support massive amounts of data transmission. As such, mmWave technology has the potential to open up many new opportunities in the fields of telecommunications, healthcare, and beyond.

One of the main opportunities offered by mmWave technology is the ability to support enhanced mobile broadband services. With its high data rates and wide bandwidth, mmWave technology is ideally suited to support the growing demand for data-intensive applications and services, such as streaming video and online gaming. This could lead to the development of new and innovative mobile services, such as virtual reality and augmented reality, which require high-speed, high-capacity communication.

Another potential opportunity offered by mmWave technology is the ability to support the Internet of Things (IoT). The IoT is a network of connected devices, sensors, and other objects that can communicate with each other and with the wider internet. With its high data rates and low latency, mmWave technology is well-suited to support the high-speed, low-latency communication required by the IoT, making it possible to connect more devices and sensors than ever before. This could lead to the development of new IoT-based applications and services, such as smart cities and connected healthcare.

In addition to these opportunities, mmWave technology also has the potential to transform the field of healthcare. With its ability to support high-speed, high-capacity communication, mmWave technology could be used to connect medical devices and sensors, enabling real-time monitoring and remote diagnostics. This could lead to the development of new and innovative healthcare applications, such as remote surgery and telemedicine, which require high-speed, low-latency communication.

Overall, mmWave technology offers many potential opportunities for the future of communication and beyond. As this technology continues to develop and expand, it has the potential to open up new and exciting possibilities in a wide range of fields, from telecommunications and healthcare to the IoT and beyond.

\subsection{Applications of Artificial Intelligence on mmWare}

Artificial intelligence (AI) and machine learning (ML) have been increasingly applied to 5G and mmWare communication in order to improve performance, reduce costs, and increase efficiency. Some examples of how AI is being used in these areas include:

\begin{enumerate}
    \item Network optimization: AI algorithms can be used to analyze network traffic and identify patterns that can be used to optimize the allocation of network resources. This can help to improve network coverage, reduce latency, and increase capacity.
    \item Predictive maintenance: AI can be used to analyze data from sensors and other sources to predict when equipment is likely to fail, allowing maintenance to be scheduled before problems occur.
    \item Self-organizing networks: AI algorithms can be used to help networks adapt to changing conditions, such as changes in traffic patterns or the addition of new devices. This can help to improve network reliability and performance.
    \item Traffic prediction: AI can be used to analyze traffic patterns and predict future demand, allowing networks to be configured to meet expected demand.
    \item Security: AI can be used to detect and respond to security threats in real-time, helping to protect networks from cyber attacks.
    \item Resource allocation: AI can be used to optimize the allocation of resources, such as bandwidth and power, in order to improve overall network performance. For example, AI algorithms can be used to predict which users are likely to require more resources at any given time and allocate those resources accordingly.
\item Network slicing: AI can be used to create and manage virtual networks, known as "slices," which can be customized to meet the specific requirements of different types of users or applications. For example, a slice might be created for a group of connected vehicles that need low latency and high reliability, or for a group of mobile users who need high-bandwidth data services.
\item Edge computing: AI can be used to process and analyze data at the edge of the network, close to the source of the data, rather than in a central location. This can reduce latency and improve the efficiency of data processing.
\item Interference management: AI can be used to predict and mitigate interference in mmWave communication systems, which can be particularly challenging due to the high frequency and limited propagation range of these signals.
\item Spectrum management: AI can be used to optimize the use of available spectrum, for example by allocating different frequencies to different users or applications based on their requirements and the conditions of the environment. This can help to improve the efficiency of the network and reduce interference.
\end{enumerate}

Overall, the use of AI in 5G and mmWare communication is still in the early stages, but it has the potential to significantly improve the performance, efficiency, and cost-effectiveness of these networks. AI can be used to optimize various aspects of network design, operation, and management, such as network optimization, predictive maintenance, self-organizing networks, traffic prediction, security, resource allocation, network slicing, edge computing, interference management, and spectrum management.

As AI technologies continue to advance and become more widely adopted, it is likely that they will play an increasingly important role in the development and deployment of 5G and mmWare communication systems. However, it is also important to consider the potential risks and challenges associated with the use of AI in these contexts, such as the need to ensure privacy, security, and ethical considerations.

\section{Discussion and Conclusion}
In a comprehensive review of the research status and development of millimeter wave communication technology,
based on the development trend, the millimeter wave communication system is analyzed as a
technical challenge. Multiplexing Gain and Diversity from Realizing Millimeter Wave Communication
Gain Angle, Analyzing Digital-Analog Hybrid Antennas for mmWave Communications
The formation of the architecture is really applied to the millimeter wave technology in the actual communication system
technical challenges and problems that still need to be solved, and introduces the current.
In implementing the IEEE 802.11aj LAN protocol specification, prototyping progress and research results obtained.

\bibliographystyle{unsrt}
\bibliography{references}

\end{document}